\documentclass[aip,apl,reprint,amssymb,amsmath,superscriptaddress,showpacs,twocolumn]{revtex4-1}
\usepackage{graphicx}
\usepackage[usenames]{color}

\begin{document}
\bibliographystyle{apsrev}

\preprint{AIP/123-QED}

\title[P. Perna \textit{et al.}]{Tailoring magnetic anisotropy in epitaxial half metallic La$_{0.7}$Sr$_{0.3}$MnO$_{3}$ thin films}

\author{P. Perna}
\thanks{Corresponding author: Dr.~Paolo Perna\\ email: paolo.perna@imdea.org}
\affiliation{Instituto Madrile\~{n}o de Estudios Avanzados en Nanociencia, IMDEA-Nanociencia, Campus Universidad Aut\'{o}noma de Madrid, 28049 Madrid,
Spain}

\author{C. Rodrigo} \affiliation{Departamento de Fisica de la Materia Condensada and Instituto "Nicol\'as Cabrera", Universidad Aut\'{o}noma de Madrid, 28049 Madrid, Spain}

\author{E. Jim\'{e}nez}
\affiliation{Departamento de Fisica de la Materia Condensada and Instituto "Nicol\'as Cabrera", Universidad Aut\'{o}noma de Madrid, 28049 Madrid, Spain}

\author{F. J. Teran}
\affiliation{Instituto Madrile\~{n}o de Estudios Avanzados en Nanociencia, IMDEA-Nanociencia, Campus Universidad Aut\'{o}noma de Madrid, 28049 Madrid,
Spain}
\author{N. Mikuszeit}
\affiliation{Instituto Madrile\~{n}o de Estudios Avanzados en Nanociencia, IMDEA-Nanociencia, Campus Universidad Aut\'{o}noma de Madrid, 28049 Madrid,
Spain}

\author{L. M\'{e}chin}
\affiliation{GREYC (UMR6072) CNRS-ENSICAEN and Universit\'{e} de Caen Basse-Normandie, Bd.~de Mar\`{e}chal Juin, 14050 Caen, France}

\author{J. Camarero}
\affiliation{Instituto Madrile\~{n}o de Estudios Avanzados en Nanociencia, IMDEA-Nanociencia, Campus Universidad Aut\'{o}noma de Madrid, 28049 Madrid,
Spain}\affiliation{Departamento de Fisica de la Materia Condensada and Instituto "Nicol\'as Cabrera", Universidad Aut\'{o}noma de Madrid, 28049 Madrid,
Spain}

\author{R. Miranda}
\affiliation{Instituto Madrile\~{n}o de Estudios Avanzados en Nanociencia, IMDEA-Nanociencia, Campus Universidad Aut\'{o}noma de Madrid, 28049 Madrid,
Spain}\affiliation{Departamento de Fisica de la Materia Condensada and Instituto "Nicol\'as Cabrera", Universidad Aut\'{o}noma de Madrid, 28049 Madrid,
Spain}

\date{\today}

\begin{abstract}
We present a detailed study on the magnetic properties, including anisotropy, reversal fields, and magnetization reversal processes, of well characterized half-metallic epitaxial La$_{0.7}$Sr$_{0.3}$MnO$_{3}$ (LSMO) thin films grown onto SrTiO$_3$ (STO) substrates with three different surface orientations, i.e.\ (001), (110) and (1$\bar{1}$8). The latter shows step edges oriented parallel to the $[110]$ (in-plane) crystallographic direction. Room temperature high resolution vectorial Kerr magnetometry measurements have been performed at different applied magnetic field directions in the whole angular range. In general, the magnetic properties of the LSMO films can be interpreted with just the uniaxial term with the anisotropy axis given by the film morphology, whereas the strength of this anisotropy depends on both structure and film thickness. In particular, LSMO films grown on nominally flat (110)-oriented STO substrates presents a well defined uniaxial anisotropy originated from the existence of elongated in-plane [001]-oriented structures, whereas LSMO films grown on nominally flat (001)-oriented STO substrates show a weak uniaxial magnetic anisotropy with the easy axis direction aligned parallel to residual substrate step edges. Elongated structures are also found for LSMO films grown on vicinal STO(001) substrates. These films present a well-defined uniaxial magnetic anisotropy with the easy axis lying along the step edges and its strength increases with the LSMO thickness.
It is remarkable that this step-induced uniaxial anisotropy has been found for LSMO films up to 120 nm thickness.
Our results are promising for engineering novel half-metallic magnetic devices that exploit tailored magnetic anisotropy.
\end{abstract}

 \maketitle

\section{INTRODUCTION}
In the past decades many improvements in the fabrication of artificial magnetic  nanostructures,\cite{miranda_2002} thin films \cite{radovic_SrO,ruotolo_2006,ruotolo_2007} and superlattices \cite{salamon_prl_2008,perna_si} have been made tailoring the properties of a large class of materials exploiting advanced techniques of patterning and stress relaxation mechanisms. For instance, it has been found that the symmetry breaking at atomic steps or anisotropic lattice relaxation are at the origin of an additional in-plane uniaxial magnetic anisotropy contribution in epitaxial magnetic thin films with cubic crystal symmetry, firstly observed in metal systems,\cite{oepen} and more recently in diluted semiconductors\cite{welp} as well as oxides.\cite{mathews_apl2005} The competition between the biaxial (four-fold) and the additional uniaxial (two-fold) anisotropy results in a magnetic reorientation, which depends on intrinsic parameters, such as substrate step density\cite{kawakami,chen,oepen,Stupakiewicz} thickness\cite{kawakami,mathews_apl2005} and angle of deposition,\cite{dijken} or extrinsic ones, such as temperature range.\cite{kawakami,mathews_apl2005,mathews2009} Hence, breaking the symmetry of magnetic systems results in additional contributions to the magnetic anisotropy, which could alter both magnetization, easy and hard axes and reversal processes.\cite{camarero_prb2008}

In order to realize \textsl{spintronics} devices such as read-heads magnetic hard disks and non-volatile magnetic memories,\cite{bowen_apl2003,park_nature} one can exploit the interesting magneto-resistive properties of the mixed-valence manganese oxides. Of particular interest is the manganite of composition La$_{0.7}$Sr$_{0.3}$MnO$_{3}$ (LSMO) showing both a Curie temperature above $300$ K and an almost full spin polarization, which can be potentially used to fabricate devices operating at room temperature (RT). The ability to control and tailor the magnetic properties of the devices is hence essential. One possibility is to engineer the growth of epitaxial films in order to obtain purpose-designed magnetic anisotropy.

It is well known that, in the case of ferromagnetic LSMO, the tensile or compressive strain induced by the film-substrate lattice mismatch can induce in-plane or out-of-plane easy magnetization directions, respectively.\cite{tsui} In particular, the strain in LSMO thin films deposited on SrTiO$_3$ (STO) (001) is in-plane tensile and an in-plane biaxial magnetic anisotropy is generally observed, with the easy in-plane direction along $\langle110\rangle$, and the hard in-plane direction along  $\langle100\rangle$.\cite{tsui,steenbeck,lecoeur,suzuki,berndt,ziese} In the case of LSMO grown onto STO(110), the substrate induces a strain that is anisotropic in-plane (i.e., the two in-plane directions of strain are inequivalent). This causes an in-plane uniaxial magnetic anisotropy with the easy axis (e.a.) of magnetization along the $[001]$ crystallographic direction.\cite{ruotolo_2006}

Another possibility to induce in-plane magnetic anisotropy is to create artificially periodic stepped surface by exploiting vicinal substrates.\cite{habermeier} These substrates are intentionally misoriented to a (near) low index surface. Step edges emerge and the high symmetry of the low index surface is broken, such that an additional uniaxial anisotropy is expected.\cite{hyman} Matthews \textit{et al.} have reported an in-plane uniaxial magnetic anisotropy at RT in $25$\;nm and $7$\;nm thick LSMO films deposited on very low miscut STO substrates ($0.13^\circ$ and $0.24^\circ$), which vanishes at low temperatures.\cite{mathews_apl2005} Uniaxial magnetic anisotropy with easy axis along the step edges has been found at $80$\;K for a $12.6$\;nm thick LSMO film deposited on a vicinal STO(001) substrate with a $10^{\circ}$ miscut off the [001] plane toward the [010] crystallographic direction.\cite{wang}

The purpose of this paper is to give a general picture on the magnetic properties, including anisotropy axis directions and magnetization reversal processes, of epitaxial LSMO thin films grown on different crystallographic directions and vicinal substrates. To do so, we first study the case of the nominally flat LSMO/STO(110), in which a well defined magnetic uniaxial anisotropy is originated from the elongated structures, which could be promoted by the anisotropic strain induced by the STO(110) surface. We compare this system with LSMO film grown onto the nominally flat STO(001) surface, which shows a weak uniaxial anisotropy probably due to residual step edges on the top film surface. Then, we demonstrate that exploiting stepped surfaces, i.e.~growing LSMO on vicinal STO(001) substrates, we can modify  artificially the film morphology and its magnetic anisotropy. In general, domain wall pinning and rotation models have been used to reproduce the angular evolution of the reversal fields, i.e., coercivity and switching fields, near the magnetization easy axis and hard axis directions, respectively.
Our results are of technological relevance in order to tailor the magnetic properties of half-metallic ferromagnetic systems.

The paper is organized as follows. In Sec.~II we describe the experimental details of the LSMO film growth, the structure, surface and magnetic characterization. The results on the structural and surface measurements are presented in Sec.~III. In Sec.~IV, we discuss extensively about the magnetic properties of the investigated systems. Finally, the summary is presented in Sec.~V concluding that the magnetic properties of epitaxial LSMO films depends strongly on the substrate induced strain, film morphology and thickness.

\section{EXPERIMENTAL}

\subsection*{Sample preparation}
The LSMO thin films were deposited by pulsed laser deposition (PLD) from a stoichiometric target onto commercially available STO(001), STO(110) and vicinal STO(001) substrates at different thicknesses (namely $16$\;nm, $70$\;nm, $120$\;nm). In the latter case the vicinal angle was $10^\circ$ from the $[001]$ surface towards the $[1\bar{1}0]$ crystallographic direction, thus inducing step edges along the $[110]$ direction (see sketches in Fig.~\ref{Fig_2}). The optimization of the growth conditions was performed on standard STO (001) substrates.\cite{perna_si} The laser fluence was  $1-2\;\text{Jcm}^{-2}$, the target-to-substrate distance was $50$\;mm, the oxygen pressure was $0.35$\;mbar and the substrate temperature was $720$\;$^\circ$C.

\subsection*{Structural and surface characterization techniques}
The crystal structure was investigated by means of X-ray Diffraction (XRD). Standard  $\theta$-2$\theta$ scans were routinely performed in order to determine the out-of-plane lattice parameters. The crystalline quality of the films was checked by measuring the Full-Width-Half-Maximum (FWHM) of the rocking curves ($\omega$-scan) and the in-plane crystal plane alignement ($\phi$-scan). XRD Reciprocal Space Mapping (RSM) around the asymmetric crystallographic peaks were also performed in order to determine the in-plane lattice parameters of the LSMO. The morphology of the samples were investigated at RT by means of atomic force (AFM) and scanning tunnel microscopies (STM), using a Nanoscope microscope. AFM and STM measurements of the film surfaces were routinely performed right after the film depositions.

\begin{table*}
\begin{center}
\resizebox{17.0cm}{!}{
\begin{tabular}[t]{c|c|c|c|c|c||c|c|c}
  % \cline{2-11}
   &  t  & \hspace{0.1 mm} in-plane \hspace{0.1 mm} & \hspace{0.1 mm} out-of-plane \hspace{0.1 mm}& $\epsilon$ $(\%$) & \hspace{0.1 mm} RMS \hspace{0.1 mm} & $M_\text{S}$ (kAm$^{-1}$) & \hspace{0.1 mm} $\mu_{0}H_\text{C}$ \hspace{0.1 mm}& \hspace{0.1 mm} $\mu_{0}H_\text{K}$ \hspace{0.1 mm}\\
  & (nm) & latt. par. (nm) & latt. par. (nm) & in-plane & (nm) & at 300 K, 0.5 T & (mT) &  (mT) \\
  \hline \hline
   LSMO(110) & 70 & 0.390$\pm$0.001 & 0.388$\pm$0.001 & $\epsilon_{[001]}$=0.8 & 1.20 & 172 & 1.50 & 12.4  \\
     & & 0.388$\pm$0.001 & &  $\epsilon_{[1\bar{1}0]}$=0.3 & & & & \\
  \hline
    LSMO(001) & 70 & 0.390$\pm$0.001 & 0.386$\pm$0.001 & $\epsilon_{[100]}$=0.8 & 0.45 & - & 0.40 & - \\
  \hline
    vicinal LSMO(001) & 16 & 0.390$\pm$0.001 & 0.386$\pm$0.001 & $\epsilon_{[100]}$=0.8 & 0.45 & 186 & 0.74 & 1.5 \\
 % \hline
   & 70 & 0.390$\pm$0.001 & 0.386$\pm$0.001 & $\epsilon_{[100]}$=0.8 & 0.79 & 200 & 0.75 & 2.5 \\
 % \hline
   & 120 & 0.390$\pm$0.001 & 0.385$\pm$0.001 & $\epsilon_{[100]}$=0.8 & 1.00 &- & 1.20 & 5.0 \\

   % \hline\hline
\end{tabular}}
\end{center}
\caption{Measured structural, morphological and magnetic parameters of LSMO films grown onto (110), (001) and vicinal (001) STO substrates for different thicknesses (\emph{t}). The in-plane and out-of-plane lattice parameters were determined by the XRD measurements (see Sec.~III).\cite{NB} Note that in the case of the (001)-oriented film the two (equivalent) in-plane axes are the [100] and [010] direction, whereas for the (110)-oriented film the two (inequivalent) in-plane axes are the [001] and $[1\bar{1}0]$ direction. The strain is defined by $\epsilon = c_0^{-1}(c-c_{0})$, where $c$ is the measured lattice parameter of the film and $c_{0}=0.387$~nm is the lattice parameter of the unstrained cubic LSMO; $\epsilon_{[001]}$ and $\epsilon_{[1\bar{1}0]}$ are the two in-plane components of the strain tensor in case of the LSMO (110)-oriented; $\epsilon_{[100]}$ is the in-plane component of the strain tensor along the $[100]$ crystallographic direction in case of the (001)-oriented LSMO. The RMS roughness is calculated from the AFM images shown in Fig.~\ref{Fig_2}. $M_\text{S}$ is the saturated magnetization extracted from the  magnetization \emph{vs.} temperature measurements, $\mu_{0}H_\text{C}$ is the coercive field and $\mu_{0}H_\text{K}$ is the anisotropy field determined by the hysteresis loops acquired at room temperature as indicated in the text (Sec.~IV).} \label{Tab_1}
\end{table*}

\subsection*{Transport and magnetic characterization techniques}
The temperature dependence of the resistivity ($\rho(T)$) of the films was measured in a 4-square contact geometry. In order to determine both, Curie temperature and saturation magnetization of the samples, magnetization versus temperature measurements were performed by using a Superconducting Quantum Interference Device (SQUID).

The magnetic anisotropy of the films and the angular dependence of the magnetization reversal were investigated at RT by high-resolution vectorial-Kerr
magnetometry measurements.\cite{camarero_prl2005} The samples were mounted in a stepper-motorized eucentric goniometer head that keeps the reflection plane fixed for the whole set of experiments. In magnetooptical measurements this is important to be able to compare the values of the magnetization components measured at different rotation angles and between different samples. In-plane vectorial-resolved hysteresis loops, i.e., $M_{\|}(H,\alpha_\text{H})$ and $M_{\bot}(H,\alpha_\text{H})$, have been acquired \emph{simultaneously} as a function of the sample in-plane angular rotation angle ($\alpha_\text{H}$), keeping fixed the external magnetic field direction. The whole angular range was probed every $4.5^\circ$, with $0.5^\circ$ angular resolution.

\section{Structural, morphological and transport characterization}

\begin{figure}[tp]
%\usepackage{graphicx}
%  \resizebox{1.0\columnwidth}{!}{%
\resizebox{8.5cm}{!}{\includegraphics[scale=1.0]{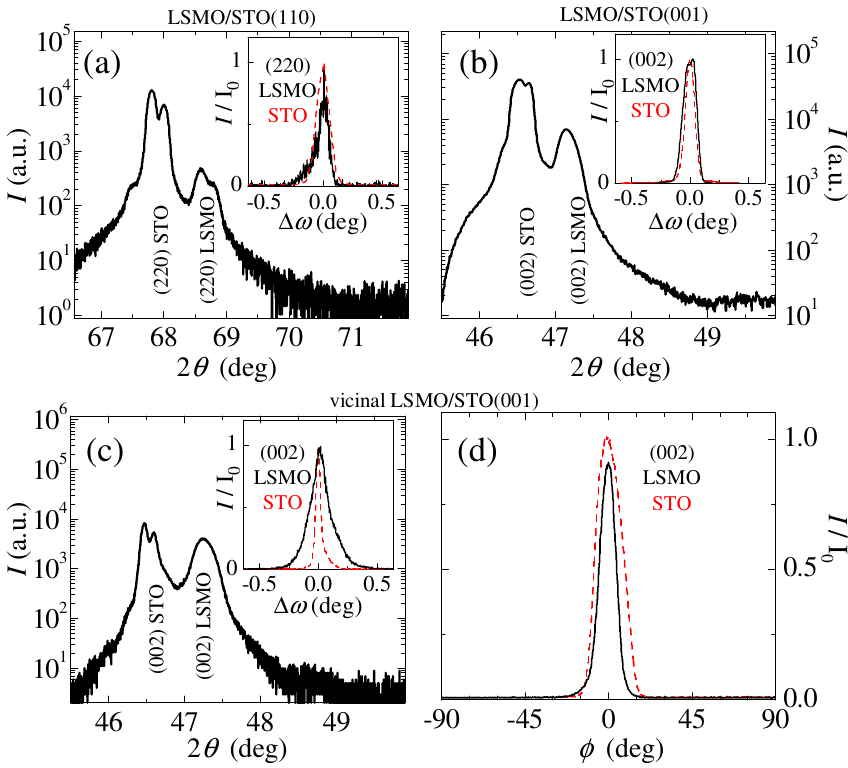}} \caption{(Color online) Structural characterization of LSMO films grown onto different STO surfaces. XRD $\theta$-2$\theta$ scans (inset $\omega$-scans) around the (220) and (002) crystallographic peaks of 70~nm thick LSMO films grown on nominally flat STO(110) (a) and STO(001) (b) substrates, and on a vicinal STO(001) (c) substrate with $10^{\circ}$ miscut off the [001] plane towards the [1$\bar{1}$0]. In the $\theta-2\theta$ scans, note that the double peak is due to the diffraction from the Cu-$K_{\alpha 1}$ and Cu-$K_{\alpha 2}$ emission lines, and that the intensity at the first substrate peak is saturated. (d) $\phi$-scan around the $(002)$ crystallographic peak of the vicinal LSMO film.} \label{Fig_1}
\end{figure}

The crystal structure of the LSMO films is determined by the STO substrate orientation. XRD $\theta-2\theta$ scans indicate that the LSMO films were epitaxially grown on the substrates (Fig.~\ref{Fig_1}). In particular, LSMO films grown onto STO(110) present the (110) axis parallel to the (110) axis of the substrate (Fig.~\ref{Fig_1}(a)), as well as LSMO films grown onto nominally flat and vicinal STO(001) present crystallographic axis collinear with those of the substrate (Fig.~\ref{Fig_1}(b)). In case of vicinal LSMO films, the offset angle was checked to be equal to the substrate vicinal angle within $\pm 0.05^\circ$ (Fig.~\ref{Fig_1}(c)). All the investigated LSMO films present high quality crystalline structure as demonstrated by the narrow rocking curves. For instance, the FHWM of the $\omega$-scans around the (220) and (002) diffraction peaks of the (110)- and (001)-oriented LSMO, respectively, were found below $0.15^\circ$ (insets of Fig.~\ref{Fig_1}). $\phi$-scan measurements performed around the crystallographic peak of all LSMO films indicate the perfect in-plane alignment of the LSMO crystal with the substrate (in Fig.~\ref{Fig_1}(d) the representative case of the LSMO grown onto the vicinal $10^{\circ}$ STO(001) substrate is shown).

\begin{figure}[tp]
  %\usepackage{graphicx}
  %\resizebox{1.0\columnwidth}{!}{%
\resizebox{8.5cm}{!}{\includegraphics[scale=1.0]{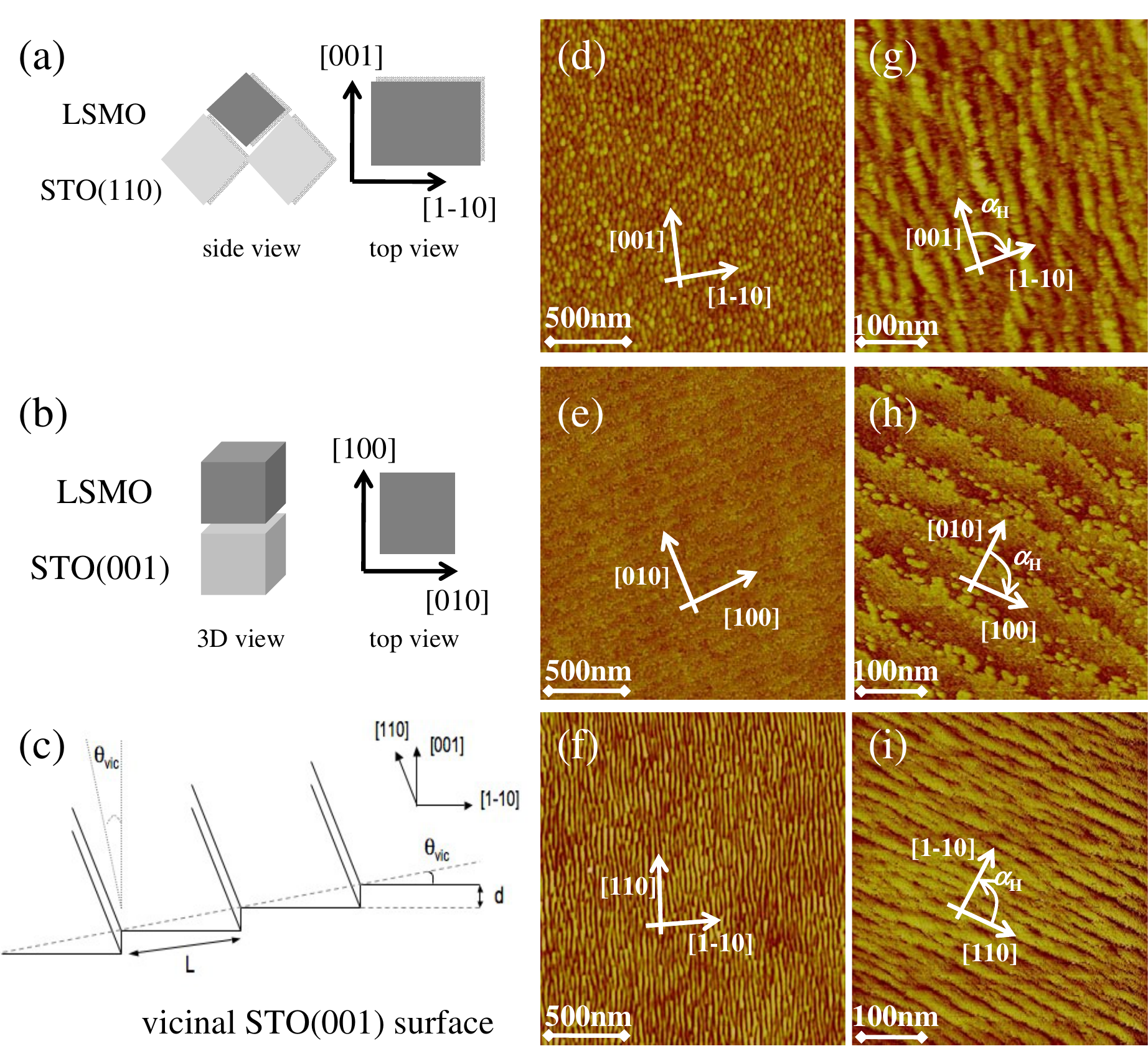}} \caption{(Color online) Morphological characterization of LSMO films grown onto different STO surfaces. Left panel: sketches of the  crystallographic cell of a LSMO film grown onto nominally flat STO(110) (a) and  STO(001) (b) substrates, and on a vicinal STO(001) substrate (c). In latter case, the vicinal STO surface is fabricated by cutting the crystal $10^{\circ}$ off the $[001]$ plane towards the $[1\bar{1}0]$, thus inducing steps along the $[110]$ crystallographic direction. Corresponding AFM images (central panel, $2\times2\;\mu\mathrm{m}^{2}$) and STM images (right panel, $500\times 500$\;nm$^{2}$) of a $70$\;nm thick LSMO film grown onto STO(110) (d-g), STO(001) (e-h) and vicinal STO(001) (f-i). Note that the morphology of the LSMO films is determined by the substrate morphology.}\label{Fig_2}
\end{figure}

The out-of-plane and the in-plane lattice parameters were determined by XRD measurements around symmetric and asymmetric crystallographic peaks.\cite{NB} The measured lattice parameters and the strain tensor along the out-of-plane and in-plane crystallographic direction for all the LSMO films are listed in Tab.~\ref{Tab_1}. Note that in the case of the (001)-oriented films the two in-plane lattice parameters of the LSMO cell are, within the error, equally tensile strained ($\epsilon_{[100]} \approx \epsilon_{[010]}$) by the substrate. Naturally, we cannot exclude minor cell distortions below experimental resolution. In contrast, in the case of the LSMO/STO(110), the two inequivalent in-plane directions of strain induced by the STO are the $[001]$ and $[1\bar{1}0]$ directions, which determine two different in-plane strain tensor components, i.e.\ $\epsilon_{[001]}$ and $\epsilon_{[1\bar{1}0]}$ (Tab.~\ref{Tab_1}).

In addition, it is worth mentioning that the measurements of temperature dependent resistivity ($\rho (T)$), performed in a 4-square contact geometry for the different LSMO films investigated, typically show very low residual resistivity ($\rho(\text{10~K})\approx 0.1\times 10^{-6}$ $\Omega$m), confirming their high crystal quality.\cite{mercone}

The morphology of the samples was investigated by means of scanning probe microscopy, in particular by AFM and STM. The average roughness (RMS) of the samples was found in the range of few unit cells (u.c.) for all samples (see Tab.~\ref{Tab_1}). In Fig.~\ref{Fig_2} we resume the morphological analysis performed on the samples. In case of the LSMO/STO(110), the particular morphology of the substrate surface induces film structures elongated along the in-plane $[001]$ crystallographic direction, corresponding to the direction of the higher in-plane tensile strain value (Fig.~\ref{Fig_2}(a,d,g), see also Tab.~\ref{Tab_1}). In case of the LSMO/STO(001), surface steps due to a small miscut of the substrate surface are found (Fig.~\ref{Fig_2}(b,e,h)). In Fig.~\ref{Fig_2}(c) we present a sketch of the vicinal surface and the typical morphology shown by LSMO films grown on it. The vicinal (1$\bar{1}$8) STO surface is fabricated by cutting the crystal $10^{\circ}$ off the [001] plane towards the [1$\bar{1}$0] and has straight atomic steps along the [110] crystallographic direction. The morphology of the LSMO films grown on such surface present structures elongated along the direction of the step edges, i.e. along the $[110]$ crystallographic direction (Fig.~\ref{Fig_2}(f,i)).

To conclude, both the crystal structure and the morphology of the LSMO films originate from induced effects by the substrate STO surfaces.

\section{Magnetic properties}
The Curie temperature ($T_\text{C}$) of the different samples is always above room temperature, as derived from magnetization versus temperature SQUID measurements (not shown), and the corresponding extracted magnetization saturation values are listed in Tab.~\ref{Tab_1}.

The study of the magnetization reversal processes and magnetic anisotropy of the films was performed at RT by measuring the in-plane vectorial-resolved hysteresis loops as a function of the in-plane angular rotation $\alpha_\text{H}$ for the whole angular range. $\alpha_\text{H}=0^{\circ}$ was referred to the crystallographic directions as labeled in Fig.~\ref{Fig_2}.

\begin{figure}[bp]
  %\usepackage{graphicx}
%  \resizebox{1.0\columnwidth}{!}{%
\resizebox{7.5 cm}{!}{\includegraphics[scale=1.0]{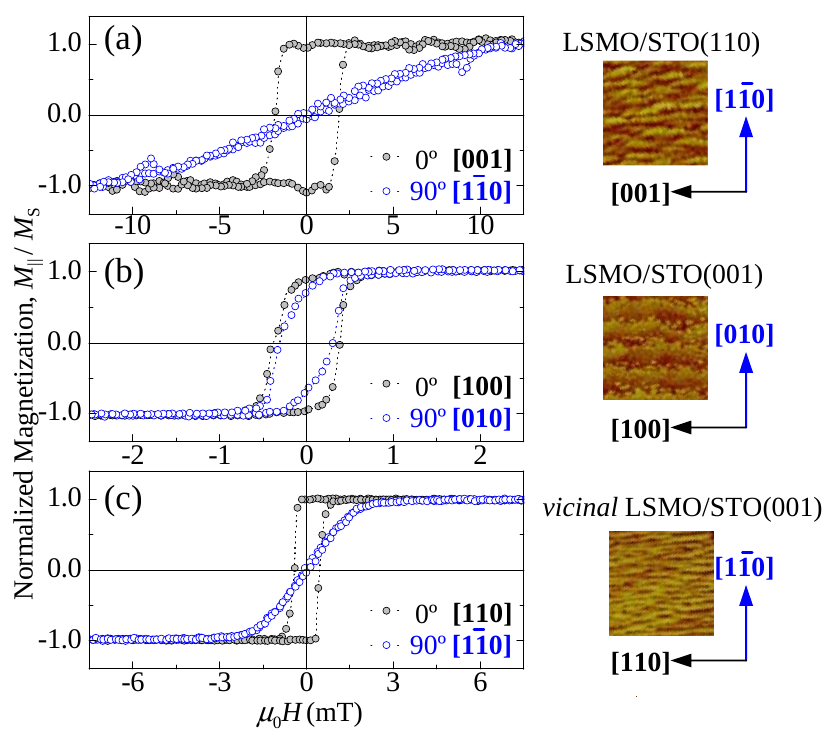}}
  \caption{(Color online) Hysteresis loops of the parallel component of the magnetization ($M_{\parallel}(H)$) at $\alpha_{\text{H}} = 0^\circ$ (filled symbols) and $90^\circ$ (open symbols), corresponding to the magnetization easy axis (e.a.) and hard axis (h.a.) directions respectively of $70$\;nm thick LSMO films grown onto STO(110) (a), nominally flat STO$(001)$ (b) and  vicinal STO (c) substrates. Note that different horizontal field scales have been used. The corresponding STM ($300\times 300\;\text{nm}^2$) images are shown to illustrate the direct connection between the LSMO film topography and their
  magnetic properties.} \label{Fig_3}
\end{figure}

To give a general view on the magnetic properties of the LSMO films grown onto different STO surfaces, we will first describe their representative hysteresis loops acquired at $\alpha_\text{H}= 0^\circ$ and $90^\circ$ with only $M_{\|}$ sensitivity (see Fig.~\ref{Fig_3}). In most experimental studies only this component is shown. In order to reliably compare them, the LSMO thickness was fixed to 70\;nm. From here, all hysteresis loops are normalized by the saturation value of the parallel component measured in the easy axis.

In general, all films present uniaxial magnetic anisotropy signatures, i.e., orthogonal easy-axis (e.a.) and hard-axis (h.a.) directions, which are related to their substrate induced morphology. In particular, a very well defined in-plane magnetic uniaxial anisotropy behavior is found for the LSMO films grown on nominally flat STO(110) (Fig.~\ref{Fig_3}(a)) and onto a vicinal STO(001) substrates (Fig.~\ref{Fig_3}(c)), i.e., films which present well defined elongated structures. This is directly reflected in the hysteresis loops acquired at $\alpha_{\rm H}=0^{\rm \circ}$ ($\alpha_{\rm H}=90^{\rm \circ}$) by remanence values of $M_{\|}$ that are close to saturation (close to zero) as well as by a maximum (zero) coercive field $\mu_{\rm 0}H_\text{C}$ when the field is oriented parallel (perpendicular) to the elongated structures. For the film grown onto a nominally flat STO(001) substrate very small differences are found when comparing the $M_{\|}(H)$ loops around $\alpha = 0^\circ$ and $90^\circ$ (Fig.~\ref{Fig_3}(b)). However, a more detailed angular dependence study of the magnetization reversal, presented below, shows the existence of a (weak) uniaxial magnetic anisotropy, which is also related to the substrate morphology, i.e., oriented parallel to residual substrate steps.

The significance of the uniaxial anisotropy of the films can be directly estimated from their corresponding loops and correlated with the topography of the films. Some specific morphological features, such as the well defined elongated structures, can be artificially achieved in LSMO films grown on both nominally flat STO(110) substrates and vicinal STO(001) substrates, as discussed above (see Fig.~\ref{Fig_2}). As a result, these films present well defined uniaxial magnetic anisotropy (Fig.~\ref{Fig_3}). In addition, the strength of this anisotropy can be derived from the magnetic field needed to saturate the films in h.a.~direction. The experimental values of the anisotropy field ($\mu_{\rm 0}H_\text{K}$) as well as the coercive field along the e.a.\ direction ($\mu_{\rm 0}H_\text{C}$) are listed in Tab.~\ref{Tab_1} and can be correlated with the topography of the films induced by the substrate.
For instance, the anisotropic in-plane strain, i.e.~$\epsilon_{[001]} \neq \epsilon_{[1\bar{1}0]}$, imposed by the STO(110) substrate gives rise to the large uniaxial magnetic anisotropy found in the LSMO(110) film.~\cite{ruotolo_2006} In turn, the existence of well oriented steps at the vicinal STO$(100)$ substrate breaks the symmetry of the LSMO$(100)$ film, in principle isotropic in-plane strained, resulting in step-induced uniaxial magnetic anisotropy.\cite{habermeier} On the other hand, larger anisotropy fields are expected for the films with larger surface roughness, as found for the LSMO film grown on the nominally flat STO(110) substrates.

The magnetization reversal mechanisms as well as a more detailed analysis of the anisotropy effects on the magnetic properties of the LSMO films grown on the different substrates cannot be understood without describing the angular dependence of the hysteresis loops of both in-plane magnetization components, i.e., $M_{\|}$ and $M_{\bot}$. This is done in detail in the following.

\subsection{LSMO grown onto nominally flat STO(110)}
\begin{figure}[bp]
  %\usepackage{graphicx}
 % \resizebox{1.0\columnwidth}{!}{%
  \resizebox{8.2cm}{!}{\includegraphics[scale=1.0]{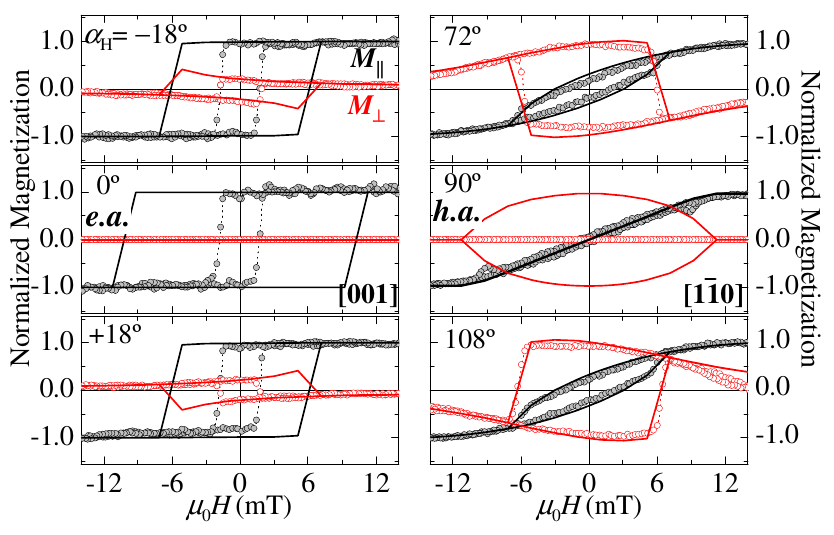}}
  \caption{(Color online) Magnetization reversal study of a $70$ nm thick LSMO film grown onto a $(110)$-oriented STO substrate around the   e.a.\ (left panel) and h.a.\ (right panel) directions. The corresponding applied field angles $\alpha_\text{H}$ are indicated in the graphs. The experimental $M_{\parallel}(H)$ and $M_{\perp}(H)$ loops are given by filled black and open red circles, respectively. The continuous lines are the corresponding simulated loops determined numerically by the Stoner-Wolfharth model without any free parameters, i.e.~by using just the uniaxial anisotropy term derived from the experimental data.}
  \label{Fig_4}
\end{figure}

Representative in-plane vectorial-resolved Kerr hysteresis loops of 70~nm thick LSMO film grown onto STO(110) acquired at selected angles $\alpha_\text{H}$ are shown in Fig.~\ref{Fig_4}. The angle $\alpha_\text{H}=0^{\circ}$ is taken when the external field is aligned parallel to the $[001]$ in-plane crystal direction, i.e.\ lying along the formed elongated structures (Fig.~\ref{Fig_2}(g)).

In general, both magnetization components, i.e., parallel ($M_{\|}$) and perpendicular ($M_{\bot}$) to the external magnetic field, show either sharp irreversible transitions or smoother fully reversible transitions. Taking into account the extended character of the film, the irreversible transitions correspond to nucleation and further propagation of magnetic domains. The reversible ones correspond to magnetization rotation processes. For $\alpha_\text{H}=0^{\circ}$ the parallel component presents a perfect squared shape hysteresis loop (central left graph of Fig.~\ref{Fig_4}). $M_{\|}$ does not change from the saturation ($M_\text{S}$) to the remanence ($M_{\|,\text{R}}$), i.e.,  $M_{\|,\text{R}}/M_{\text{S}}\approx 1$, and there is only a sharp irreversible jump at the coercive field $\mu_\text{0}H_\text{C}=1.50$~mT, in which the magnetization reverses completely. In turn the perpendicular component is negligible in the whole field loop, i.e., $M_{\bot}(H)\approx 0$. Both are expected behaviors of a magnetization e.a.~direction, in which the magnetization reversal takes place via nucleation and further propagation of magnetic domains oriented parallel to the field direction.

For $\alpha_\text{H}\neq0^{\circ}$, clear $M_{\bot}(H)$ loops with both reversible and irreversible transitions are found, in correspondence to the $M_{\|}(H)$ loops, as shown in the top and bottom graphs of Fig.~\ref{Fig_4}. In particular, for $\alpha_\text{H}\pm 18^{\circ}$ the irreversible switching field of the perpendicular component is $\mu_\text{0}H_\text{S}(\pm 18^{\circ})=1.55$~mT, identical to $\mu_\text{0}H_\text{C}(\pm 18^{\circ})$. In addition, the $M_{\bot}(H)$ loops acquired at opposite angles present similar shape but different sign. The latter arises from the sensitivity of $M_{\bot}(H)$ to the anisotropy direction.~\cite{erika_prb2009} Therefore, around the e.a.~direction, the reversible transitions correspond to a reversal by magnetization rotation whereas the irreversible ones correspond to  propagation of magnetic domains not oriented parallel to the field direction but to the e.a.~direction.

\begin{figure}[tp]
  %\usepackage{graphicx}
%  \resizebox{0.80\columnwidth}{!}{%
\resizebox{8.0cm}{!}{\includegraphics[scale=1.0]{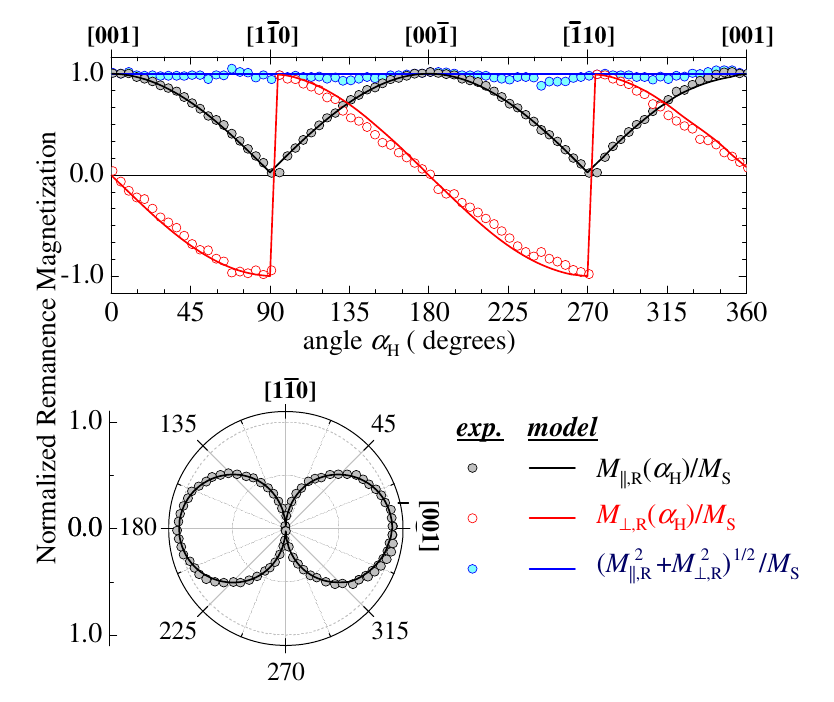}}
  \caption{(Color online) Angular dependence of the normalized remanence magnetization ($M_{\|,\text{R}}/M_\text{S}$ and $M_{\bot,\text{R}}/M_\text{S}$) of a $70$ nm thick LSMO film grown onto a nominally flat $(110)$-oriented STO substrate. The inset (bottom graph) shows the polar-plot representation of $M_{\|,\text{R}}$. The symbols (solid lines) are the experimental (predicted) values derived from the vectorial-resolved Kerr measurements (numerical simulations) as those shown in Fig.~\ref{Fig_4}.}
\label{Fig_5}
\end{figure}

When the field is applied perpendicular to the elongated structures, i.e., $\alpha_\text{H}=90^{\circ}$, the $M_{\|}(H)$ loop shows an almost linear and
reversible behavior of the magnetization, $M_{\|,\text{R}}/M_{\text{S}}\approx 0$, and $\mu_\text{0}H_\text{C}\approx 0$~mT (see central right graph of Fig.~\ref{Fig_4}). This features are typical of an uniaxial magnetic anisotropy hard-axis. In this case, the anisotropy field extracted from the loop is $\mu_\text{0}H_\text{K}= 10$~mT. The negligible signal found in the corresponding $M_{\bot}(H)$ loop is related with the perfect alignment of the external field with the h.a.~direction and the acquisition procedure. In particular, a vanishing perpendicular component turns out after averaging many successive iterations, in which for each one the magnetization would rotate alternatively along the positive and negative values of $M_{\bot}$. This is confirmed by the hysteresis loops acquired around the h.a.~direction, e.g., $\alpha_\text{H} = 90^{\circ}\pm 18^{\circ}$, which show similar (and large) $M_{\bot}$ signals but with opposite sign. Additionally, around the h.a.~direction, it can be seen that $\mu_\text{0}H_\text{S}(90^{\circ}\pm 18^{\circ})>\mu_\text{0}H_\text{C}(90^{\circ}\pm 18^{\circ})$. Therefore, the magnetization reversal close to the h.a.~direction is governed by rotation processes where the magnetization tries to be aligned parallel to the e.a., i.e., towards  the elongated structures.

\begin{figure}[tp]
\resizebox{8.0cm}{!}{\includegraphics[scale=1.0]{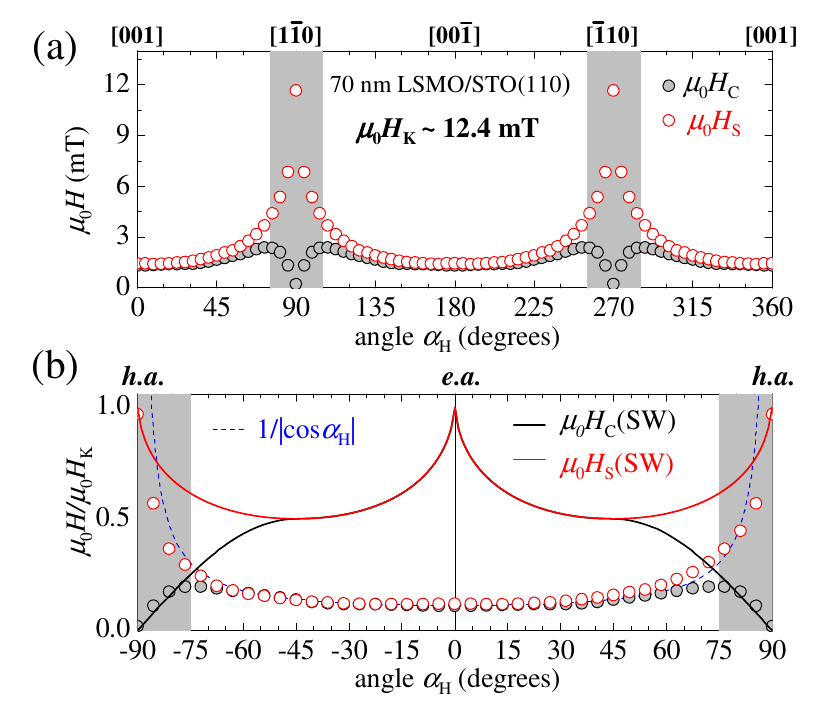}}
  \caption{(Color online) (a) Angular dependence of the coercive field $\mu_{0}H_{\text{C}}$ and switching field $\mu_{0}H_{\text{S}}$ of a $70$\;nm thick LSMO film grown onto a nominally flat $(110)$-oriented STO substrate extracted from the vectorial-resolved Kerr measurements, as those shown in Fig.~\ref{Fig_4}. (b) Comparison of the experimental data (symbols) with the predicted behavior derived from the rotation model (solid lines) and the pinning model (dashed line). The shadowed areas indicate the angular range where reversible rotation processes are the relevant mechanism during reversal.} \label{Fig_6}
\end{figure}

For a more quantitative analysis, considerable magnetic parameters such as remanence magnetization, coercivity, and switching field, have been readily obtained as a function of the angle $\alpha_\text{H}$ from the hysteresis loops like those shown in Fig.~\ref{Fig_4}. In general, the uniaxial anisotropy of the film is clearly observed in the consequent angular plots (see Fig.~\ref{Fig_5} and Fig.~\ref{Fig_6}).

The angular dependence of the normalized remanence values of both magnetization components, i.e., $M_{\|,\text{R}}/M_{\text{S}}$ and $M_{\bot,\text{R}}/M_{\text{S}}$, is shown in Fig.~\ref{Fig_5}. There is a pronounced oscillation of both magnetization components with periodicity of 180$^{\circ}$, the parallel component follow a $|\cos\alpha_\text{H}|$ law dependence, the perpendicular component changes the sign when a characteristic direction, i.e., e.a.\ and h.a.\ directions, is crossed, and both components are complementary, i.e.,  $M^2_\text{S}=(M^2_{\|,\text{R}}+M^2_{\bot,\text{R}})$. The polar-plot of $M_{\|,\text{R}}/M_{\text{S}}$ shown in the inset of Fig.~\ref{Fig_5} shows the
characteristic "two-lobe" behavior originated from a two-fold magnetic symmetry. All these features confirm the uniaxial magnetic anisotropy behavior of the film, where the anisotropy axis is aligned parallel with the direction of the elongated grains, i.e.\ parallel to the in-plane $[001]$ crystallographic direction (see Fig.~\ref{Fig_2}). The data have been properly reproduced in the whole angular range with the coherent rotation Stoner-Wohlfarth model\cite{stoner} by using the uniaxial anisotropy field found experimentally. To note that, in order to reproduce satisfactorily the experimental data, the biaxial anisotropy term has to be neglected, which confirm that at remanence the magnetization is aligned along the anisotropy axis.

The two-fold symmetry of the magnetic properties is also found in the experimental data of the angular dependence of both coercive ($\mu_{0}H_{\text{C}}$) and switching ($\mu_{0}H_{\text{S}}$) fields, as revealed the by the $180^\circ$ periodicity of both properties (Fig.~\ref{Fig_6}(a)). Two angular ranges can be defined. In a wide angular region around the e.a.\ direction, i.e., $|\alpha_\text{H}|<75^{\circ}$, both fields are similar and follows a $1/|\cos\alpha_\text{H}|$ law (discontinuous line in Fig.~\ref{Fig_6}(b)), accordingly to the domain pinning model  prediction,\cite{chikazumi} which includes reversal generated by pinned domain wall propagation processes. This has been already observed in both  perpendicular\cite{givordjmmm98} and in-plane anisotropy systems.\cite{prieto_IEEE2008} Thus, nucleation and further propagation of pinned magnetic domains is the relevant process during the irreversible transitions.

Close to the h.a.~direction, i.e., $|\alpha_\text{H}|>75^{\circ}$, the pinning model cannot reproduce the experimental data, whereas the rotation model can do so satisfactory. For instance, the coercive field (switching field) decreases (increases) to zero (up to the anisotropy field) as approaching the h.a.\ direction, as predicted by the rotation model (solid lines in Fig.~\ref{Fig_6}). This indicates that the magnetization reversal is governed mainly by rotation processes close to the h.a. direction. Note that the rotation model only reproduces their angular evolution around the h.a.~directions, where reversible processes are the relevant mechanism during reversal (filled area in Fig.~\ref{Fig_6}). It fails around the e.a. directions, where irreversible (nucleation and propagation of magnetic domains) processes dominate, as described above.

Finally, it is worth to mention that the rotation model has also been used to extract the theoretical $M_{\parallel}(H)$ and $M_{\perp}(H)$ curves in the whole angular range and, apart of the overestimation of the reversal fields around the e.a. directions, the model reproduce satisfactorily the experimental data, as shown the continuous lines in Fig.~\ref{Fig_4}.

Concluding, the experimental values of the anisotropy field  ($\mu_{0}H_\text{K}$) as well as the coercive field along the e.a.~direction ($\mu_{0}H_\text{C}$) (listed in Tab.~\ref{Tab_1}) can be correlated with the topography of the films, which might be induced by the substrate strain.

\subsection{LSMO grown onto nominally flat STO(001)}
\begin{figure}[bp]
%\usepackage{graphicx}
% \resizebox{1.0\columnwidth}{!}{%
\resizebox{8.0cm}{!}{\includegraphics[scale=1.0]{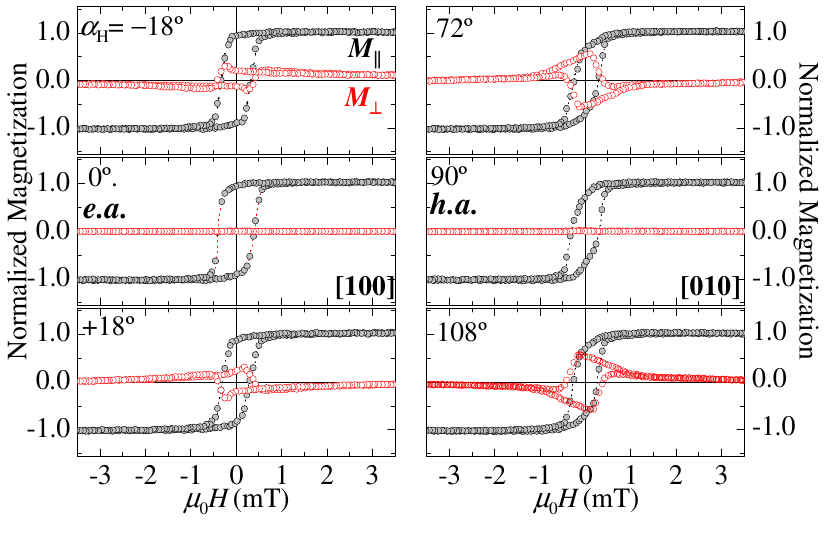}} \caption{ (Color online) Magnetization reversal study of the $70$ nm thick LSMO film grown onto nominally flat $(001)$-oriented STO substrate around the e.a.~(left panel) and h.a.~(right panel) directions. The corresponding applied field angles $\alpha_\text{H}$ are indicated in the graphs. The experimental $M_{\|}(H)$ and $M_{\bot}(H)$ loops are given by filled black and open red symbols, respectively. Notice the change of sign of the $M_{\bot}(H)$ loop when the characteristic axes are crossed.} \label{Fig_7}
\end{figure}

A different magnetization reversal behavior is found in the LSMO film grown on a nominally flat STO(001) substrate. Fig.~\ref{Fig_7} shows representative Kerr hysteresis loops of a $70$~nm thick LSMO film grown onto STO(001) acquired at selected angles $\alpha_\text{H}$. In this case, $\alpha_\text{H}=0^{\circ}$ is taken when the external field is aligned parallel to the $[100]$ in-plane crystal direction, i.e., parallel to the substrate step direction. The angular evolution of the parallel component of the magnetization ($M_{\|}(H)$) reveals only very small changes when comparing the behavior around $\alpha_\text{H}=0^\circ$ and $90^\circ$, suggesting an almost isotropic behavior, as  commented above.

However, by looking for the change of sign of the $M_{\perp}$ loops when a characteristic direction is crossed (Fig.~\ref{Fig_7}), we are able to precisely locate both easy and hard axis directions, as revealed the evolution of the left and right graphs, respectively. In particular, the $M_{\bot}(H)$ loop vanishes progressively when approaching $\alpha_\text{H}=0^{\circ}$ from negative angles changing its sign for positive ones, whereas it suddenly changes around $\alpha_\text{H}=90^{\circ}$. These features are expected for the characteristic e.a.~and~h.a.~directions of uniaxial magnetic anisotropy systems.
Therefore, the e.a.~direction is parallel to the $[100]$ crystallographic direction and lies along the surface steps, and the h.a.~direction is perpendicular to it. On the contrary, the h.a.~loop presents non zero coercivity (it is only slightly smaller than the coercivity of the e.a.) and relevant irreversible processes. These features are expected in not well defined uniaxial anisotropy systems where magnetic domains with many different magnetic orientations can be nucleated during reversal. For instance, in this particular case, the very low roughness of the film (see Fig.~\ref{Fig_2}(e,h)) could explain the small effects of the uniaxial anisotropy. In addition, a nonnegligible biaxial magnetocrystalline anisotropy could not be excluded, as found in Ref.~\cite{perna_jap11} in LSMO grown onto $2^{\circ}$ miscut STO(100). In short, the LSMO(100) films grown on nominally flat STO(100) surfaces present a weak uniaxial anisotropy at RT. This is aligned parallel to the $[100]$ crystallographic direction, i.e., parallel to the substrate step edges, which suggests that it is induced by these surface steps.

\subsection{LSMO grown onto vicinal STO(001)}
\begin{figure}[bp]
\resizebox{7 cm}{!}{\includegraphics[scale=1.0]{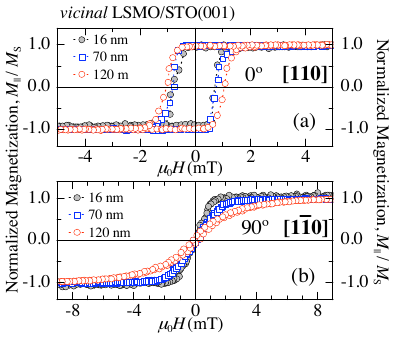}} \caption{ (Color online) Hysteresis loops of the parallel component of the magnetization ($M_{\parallel}(H)$) acquired at the e.a.~(a) and h.a.~(b) directions of $16$ nm, $70$ nm and $120$ nm thick LSMO films grown onto vicinal STO(001) substrates. The horizontal axes have been scaled differently. Notice that the anisotropy field increases as the thickness increases.} \label{Fig_8}
\end{figure}

The study of the anisotropy and magnetization reversal of LSMO films grown on vicinal STO(001) surfaces has been performed for different film thicknesses, from $16$\;nm up to $120$\;nm. Fig.~\ref{Fig_8} provides a general view on the influence of the LSMO thickness on the magnetic anisotropy. $\alpha_\text{H}=0^{\circ}$ is taken when the external field is aligned parallel to the substrate step edge direction, i.e.~along the $[110]$ crystallographic direction.

\begin{figure}[bp]
  %\usepackage{graphicx}
  %\resizebox{1.0\columnwidth}{!}{%
\resizebox{8.0cm}{!}{\includegraphics[scale=1.0]{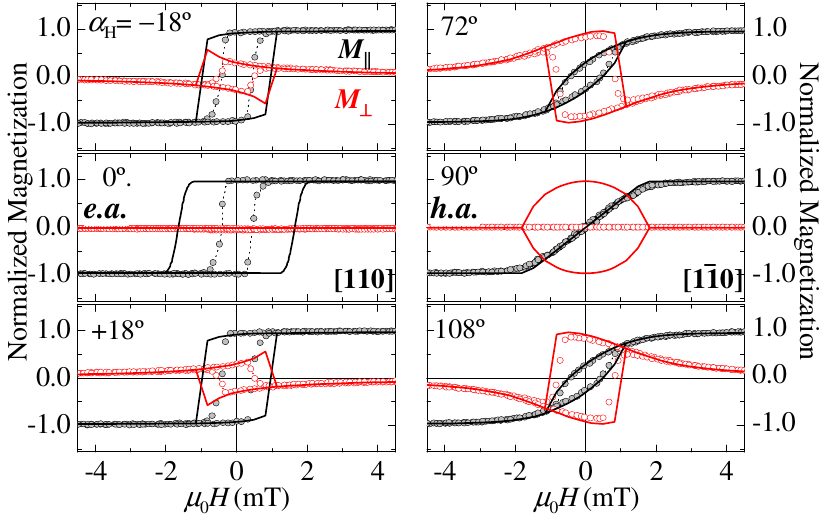}} \caption{ (Color online) Magnetization reversal study of the $70$ nm thick LSMO film grown onto vicinal $10^\circ$ STO(001) substrate around the e.a.~(left panel) and h.a.~(right panel) directions. The corresponding applied field angles $\alpha_\text{H}$ are indicated in the graphs. The experimental $M_{\parallel}(H)$ and $M_{\perp}(H)$ loops are given by filled black and open red symbols, respectively. The solid lines represent the simulated loops obtained by using a coherent model described in Appendix, which only considers the uniaxial anisotropic term derived from the experimental data.}\label{Fig_9}
\end{figure}

Fig.~\ref{Fig_9} shows representative Kerr hysteresis loops of a $70$~nm thick LSMO film grown onto vicinal STO(001) surface, with a $10^{\circ}$ miscut off the $[001]$ plane toward the $[1\bar{1}0]$ crystallographic direction, at selected angles $\alpha_\text{H}$ between the magnetic field and the surface steps direction. Again, $\alpha_\text{H}=~0^{\circ}$ is taken when the external field is aligned parallel to the $[110]$ in-plane crystal direction. As before, the characteristic axes are located precisely at the change of sign of the $M_{\perp}(H)$ loops. Hence, the e.a.\ direction is along the direction of the steps (i.e., $[110]$) and the h.a.\ is perpendicular to it (i.e., $[1\bar{1}0]$). In comparison with the film grown onto a nominally flat STO(001) surface, the LSMO films grown onto the vicinal STO(001) present a well defined uniaxial anisotropy behavior, originating from the periodic stepped substrate structure, which induces the formation of well oriented elongated structures along the step edge direction.

For all samples, a well defined uniaxial magnetic anisotropy is found, as shown by the e.a.~($0^{\circ}$) and h.a.~($90^{\circ}$) $M_{\|}(H)$ loops in Fig.~\ref{Fig_9}. In all cases the anisotropy is oriented parallel to the substrate step edged, i.e., parallel to the elongated structures. In addition, from the anisotropy field values extracted from the h.a.~loops (see Tab.~\ref{Tab_1}), larger anisotropy fields, i.e., stronger uniaxial anisotropy, are found when the LSMO thickness increases. Remarkable is that this step-induced uniaxial anisotropy has been found for all LSMO films up to $120$\;nm thickness.

The evolution of the vectorial-resolved hysteresis loops is analogous to the flat LSMO(110) film. Similarly, the numerical simulations performed with the coherent model just by considering the uniaxial term extracted from the experiments, reproduce qualitatively the experimental vectorial-resolved hysteresis loops for the whole angular range (see solid lines in Fig.~\ref{Fig_9}), but overestimate the reversal fields close to the e.a.~direction. As a consequence the same arguments and conclusions can be derived for the vicinal LSMO(001) film. The relevant mechanism for the magnetization reversal close to the e.a.~direction is the nucleation and propagation of magnetic domains oriented parallel to the elongated structures, i.e., parallel to the steps, whereas reversal by rotation processes are the relevant mechanism close to the h.a.~direction, i.e., perpendicular to the steps. Furthermore, for this particular sample, this picture has been strongly supported recently in real space by means of angular dependence Kerr microscopy  measurements.~\cite{perna_NJP}
For all thicknesses investigated, the uniaxial anisotropy of the films is clearly observed in the angular plots of the remanence magnetization extracted from the experimental vectorial-resolved hysteresis loops acquired in the whole angular range (similar to that of Fig.~\ref{Fig_5}, not shown), presenting 180$^{\circ}$ periodicity.

\begin{figure} [tp]
  %\usepackage{graphicx}
%  \resizebox{0.80\columnwidth}{!}{%
\resizebox{7.0 cm}{!}{\includegraphics[scale=1.0]{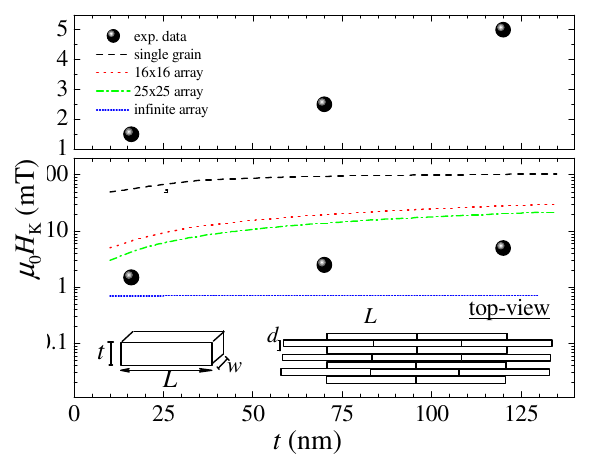}} \caption{(Color online) Thickness-dependent anisotropy field ($\mu_{0}H_\text{K}$) derived from the experimental hysteresis loops shown in Fig.~\ref{Fig_9} (top graph). Bottom graph shows the comparison between the experimental data (symbols) and the numerical simulations described in the text which consider magnetostatic interaction between interacting magnetic grains (cuboids-like) within different arrays (solid lines). The dimensions of the cuboid are taken from the parameters obtained from the AFM images: length $L = 1\;\mu$m, wide $w = 30$\;nm, thickness $t$ and intergrain distance $d = w + 0.2$\;nm. Insets show the sketch of a single grain and an array of $n \times m$ interacting grains, respectively.} \label{Fig_10}
\end{figure}

Finally, in order to understand the rising of the anisotropy field with the film thickness  (Tab.~\ref{Tab_1}), we have used a simple model that considers the films like a collection of physically independent magnetic grains with magnetostatic interactions.\cite{tsymbal} This ansatz is required as we can exclude change of the anisotropy due to inhomogeneous strain release within our experimental resolution. The magnetic grains are approximated by cuboids of dimensions $L,w,t$, where $L = 1$~$\mu\text{m}$, $w = 30$~nm are the measured values extracted from the AFM images (Fig.~\ref{Fig_2}) and $t$ is assumed to be the film thickness (Tab.~\ref{Tab_1}). The gap between the cuboids has been considered as half u.c., i.e., $0.2$\;nm. Experimentally the anisotropy field presents a monotonous increase of $\mu_{0}H_\text{K}$ with the LSMO thickness, as shown Fig.~\ref{Fig_10}(a). In Fig.~\ref{Fig_10}(b) we compare the experimental data (black symbols) with the simulated behavior for a single grain (black dashed line, sketch in Fig.~\ref{Fig_10}(c)), for an array of $16 \times 16$ and $25 \times 25$ grains (red dashed and green dashed-dotted lines, sketch in Fig.~\ref{Fig_10}(d), respectively) and for an infinite array (blue solid line). It is clear that the limits of the simulation, i.e., single grain and infinity array, are far from the measured values, but in between the model predicts an increasing of the anisotropy field with the thickness, as observed experimentally.

\section{CONCLUSIONS}
X-Ray diffraction measurements proved the high crystalline quality and the epitaxy of our LSMO films, which always show a Curie temperature above room temperature. Scanning probe microscopies were used in order to investigate the morphology of the samples. LSMO films, with thicknesses ranging from $16$ nm to $120$ nm, show very low surface roughness (in the unit cells range) whereas the morphology is induced by the STO substrate.

We have investigated the magnetic properties of three different systems based on epitaxial LSMO thin films by vectorial-resolved magneto optical Kerr magnetometry. LSMO films deposited onto (110)-oriented STO surfaces show a well defined uniaxial magnetic anisotropy ascribed to the substrate induced strain. In this case the two in-plane orthogonal directions are not equivalent, and elongated structures parallel to the [001] crystallographic direction are found. In the case of the LSMO films grown onto nominally flat STO(001) we found a weak uniaxial magnetic anisotropy that we ascribed to the small miscut angle of the substrate surface. Finally, LSMO films, up to $120$\;nm thickness, grown on vicinal STO substrates, present 2-dimensional elongated structures as well, running parallel to the substrate step edge direction. In such a system, the in-plane steps along the [110] crystallographic direction, imposed by the $10^{\circ}$ vicinal cut of the STO substrate, cause a well defined uniaxial magnetic anisotropy. In this case, the easy axis for the magnetization lies parallel to the step edges, i.e. parallel to the [110] crystallographic direction, whereas the $[1\bar{1}0]$,  perpendicular to the steps, is the hard axis for the magnetization.

For all systems investigated, we have shown the angular dependence of the magnetization reversal processes from the detailed analysis of the vectorial-resolved Kerr loops. Nucleation and further propagation of magnetic domains and rotation processes are the relevant mechanism during magnetization reversal for $\alpha_{\rm H}$ near the e.a.~and h.a.~directions, respectively. The angular dependence of the reversal fields, both switching and coercive fields, around the e.a.~and h.a.~directions has been understood in the framework of the pinning and rotation models, respectively. In addition, the latter becomes more relevant for the thinnest LSMO films.
Finally, we have shown that the rising of the induced anisotropy with the film thickness behavior can be reproduced qualitatively using a simple model based on magnetostatic interactions only.

In conclusion, we have shown that the magnetic properties of LSMO epitaxial thin films, grown in different crystallographic directions and on vicinal substrates, strongly depend on the substrate induced effects. Remarkable that in the case of the step-induced uniaxial anisotropy by using vicinal STO(100) surfaces, a well defined uniaxial anisotropy has been found for LSMO films up to 120 nm thickness. The ability to control and tailor the magnetic properties of LSMO thin films results thus to be an important task for the design of novel devices based on thin film technology.

\section*{ACKNOWLEDGEMENTS}
This work was supported in part by the Spanish MICINN through Projects No.~CSD2007-00010 and MAT2010-21822, and by the Comunidad de Madrid through Project
No.~S2009/MAT-1726. P.P.~thanks the European Science Foundation (ESF) through the activity entitled 'Thin Films for Novel Oxide Devices'
(http://www.ims.tnw.utwente.nl/thiox/) for partial financial support through exchange grants. The authors wish to thank Ch.~Simon at CRISMAT (UMR6508) for
SQUID magnetization measurements and CRISMAT for the LSMO target fabrication. The authors are also very grateful to U.~Scotti di Uccio at CNR-SPIN Naples
for fruitful discussions and for the use of the X-ray diffractometer.


\begin{thebibliography}{36}
\bibitem{NB}
The crystal structure of the films was determined by performing XRD lattice mappings around symmetrical and asymmetrical diffraction peaks. For more
details see for example Ref. \cite{perna_si}.
\end{thebibliography}
\end{document}